# A Liquid Democracy System for Human-Computer Societies

**Anton Kolonin, Ben Goertzel, Cassio Pennachin, Deborah Duong, Marco Argentieri, Matt Iklé, Nejc Znidar**

SingularityNET Foundation, Amsterdam, Netherlands
{anton, ben, cassio}@singularitynet.io

## Abstract

Problem of reliable democratic governance is critical for survival of any community, and it will be critical for communities powered with Artificial Intelligence (AI) systems upon developments of the latter. Apparently, it will be getting more and more critical because of increasing speeds and scales of electronic communications and decreasing latencies in system responses. In order to address this need, we present design and implementation of a reputation system supporting "liquid democracy" principle. The system is based on "weighted liquid rank" algorithm employing different sorts of explicit and implicit ratings being exchanged by members of the society as well as implicit assessments of of the members based on measures of their activity in the society. The system is evaluated against live social network data with help of simulation modeling for an online marketplace case.

## 1 Introduction

Over the history of human communities, no reliable form of reaching truly long-term democratic consensus has been invented [Hazin and Shheglov, 2018]. While different sorts of social organizations have been tried, starting with completely centralized governance in from of "monarchy" and ending with completely distributed "anarchy" with all sorts of "democracy" in-between, any sorts of "democratic" governance requires some sort of consensus ground to be recognized and accepted by entire society.

The first form of consensus is known to rely on brute force in animal groups and ancient human societies and it can be serving to the minority having the access to the force – this problem is replicated in modern distributed computing system based on Proof-of-Work (PoW) in blockchain environments. More advanced form of consensus mostly usable by human race nowadays is reached on basis of financial capabilities of members of community, and it is known to lead the situation when "reacher become richer" and gain more and more power – this problem is also replicated in latest developments of distributed computing system based on Proof-of-Stake (PoS) in blockchain. In the blockchain systems, including some that are now used to design ecosystems for Artificial Intelligence (AI) applications, the mostly suggested solution called Delegated Proof-of-Stake (DPoS), which effectively means that rule on basis of financial capabilities while it is implemented indirectly, by means of manually controlled voting process to select delegates to conduct the governance of the system. The latter be only limited improvement and can nor be implemented in AI communities operating at high speeds not controllable by means of limited human capabilities.

The described situation leads to the danger that consensus in any emergent AI community may be quickly took over by AI system hostile to either majority of community of AI systems or humans that are supposed to be served by given AI community. It may be either because of emergent hostility of an AI system in respect to humans or because of particular of people managing given AI system in favor of human minority damaging majority.

## 2 Suggested Solution

To solve the problem, we suggest future distributed systems involving social interactions between humans as wells as AI systems to be based on Reputation Consensus implementing Proof-of-Reputation (PoR) principle, opposing power of brute force (PoW), power of money (PoS or DPoS). The Proof-of-Reputation can make it possible to implement system of Liquid Democracy to fix known forms of representative democracy, influenced by power of money as we know in human history. It can be helpful for direct democracy as well, enabling reliable implementation of the latter at scale of modern real-world communities, human or artificial.

The Reputation Consensus principle asserts that governing power of member of a human or artificial society depends on Reputation of the member earned on basis of the following principles, as shown on Figure 1.

The first key principle is that Reputation may be computed by means of different measures, called "ratings", performed explicitly or implicitly by all members of community, called "raters", in respect to ones who Reputation is being computed for, called "ratees", with account to Reputations of the "raters" themselves.

The second key principle is time scoping of the Reputation computation, so that measures collected by a ratee in the past are less contributing to its current Reputation than the latest ones, which have more impact.

The third key principle if openness of all Reputations of all raters and ratees with the ratings that they issue or receive so that audit of Reputations and the historical measures over the history can be performed in order to prevent Reputation cheating and gaming.

The fourth key principle is precedence of human measures over artificial ones, so that ratings provided by human participants of a hybrid human-machine communities have unconditional precedence over the ratings provided by AI systems, if they are also capable to contribute to evaluation of humans and artificial entities in a community.

Multiple kinds of explicit and implicit measures contributing to evaluation of Reputation may be considered, depending on implementation of a given Reputation system. Applicability of the measures or ratings may depend on accuracy and reliability that they may provide as well as resistance to attack vectors targeting takeover of the consensus by means of reputation cheating and gaming. Primarily, we consider such measures as: a) members explicitly staking financial values on other members; b) members explicitly providing ratings in respect to transactions committed with other members; c) implicit ratings computed from the financial values of transactions between the members; d) evaluation of textual, audial and video reviews or mentions made by members in respect to other members or transactions between them.

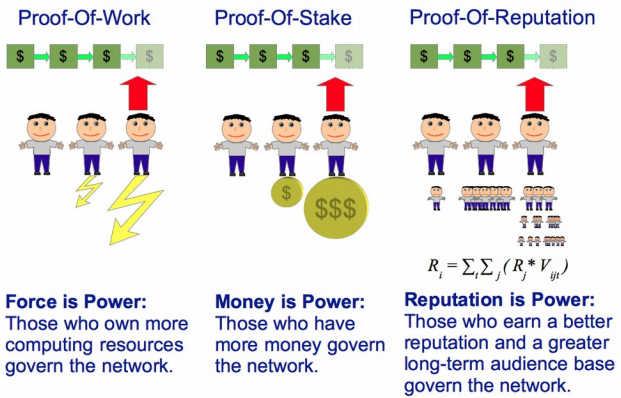

Figure 1. Types of consensus in distributed systems such as Proof-of-Work, Proof-of-Stake and Proof-of-Reputation.

Should be noted that we consider the problem of Consensus for distributed communities primarily because centralized and even decentralized solutions of AI systems may simply become targeting interests of limited group of powerful and resourceful humans having full control, and hence unlikely be serving interests of entire community as a whole.

Our goal has bee set to creation of experimental Reputation System performing Reputation calculations ad supplying distributed Reputation Consensus for any artificial, human or hybrid communities, including the SingularityNET ecosystem is one of the first practical cases. So out study has began with building formal mathematical modeling and multi-agent simulations in order to figure out the best set of measures and settings resistant to possible consensus takeover attack vectors. Multiple design solutions for reputation systems serving the purpose has been found, such as discussed by other authors: [Swamynathan *et al*., 2010], [Sänger and Pernul, 2018]. The most critical part of any system intended for this purpose appears to be the ability to handle a high degree of anonymity [Androulaki *et al*., 2008] of market participants, as it is characteristic of modern distributed [Gupta *et al*., 2003] ecosystems, including ones that are based on public networks [Blömer *et al*., 2015] such as blockchains.

## 2 Reputation System Implementation

The expected objective of the reputation system is to compute reliable, trustable and consistent reputation rankings for every participant of any kind of society providing reputation consensus to build any kind of democratic governance in a community of any kind. At the same time, specific communities may be employing different configurations of the same reputation algorithm.

Initially, given the requirements above and the prior art, we have ended up with a reputation system design based on the "**weighted liquid rank**" (WLR) algorithm [Kolonin *et al*., 2018] applied as a generic computational framework serving different applications.

For the starting point for the WLR concept development we took the "**page rank**" idea [Brin and Page, 1998], which has been extended with variety of different sorts of ratings and account for temporal scoping with notion of reputation decay. Later on, the implementation of the Reputation System has been built into open source Aigents project in Java (https://github.com/aigents/aigents-java) and applied to study open social network data on Steemit blockchain platform [Kolonin, 2019].

Further on, the open source implementation of the entire W L R  a l g o r i t h m  h a s  b e e n  d o n e  i n  P y t h o n (https://github.com/singnet/reputation) based on the need for reputation assessment in marketplaces where consumers and not necessarily suppliers at the same time so the raters are not necessarily ratees themselves and the **page rank** idea can't be applied. It has been explored how it can be used to distinguish honest and dishonest market participants under different market conditions defined by transactional patters with different transaction frequencies and costs [Kolonin *et al.*, 2019].

In the very recent work, we have been exploring the market conditions where the honest and dishonest market participants have the same transactional patterns in terms of frequencies and costs so they can not be distinguished on that basis. Under these conditions, the suppliers were still not overlapping with consumers so **page rank** won't work either. In order to solve the problem, we have implemented extension for the WLR algorithm so the reputation of the raters can be implicitly assessed by the time they spend on the market (Time-on-the-Market or TOM) or the spendings that they make on this market (Spending-on-the-Market or SOM). Hence, two following extensions to the reputation system has been made.

1) **TOM-based** reputation system: In addition to weighting ratings with financial values per-transaction, weights the ratings based on the rater's time on the market (TOM) as a "proof-of-time". That is, the raters (buyers) are implicitly rated based on how long they have been on the market. So, rating by buyer with a longer history influences reputation of a seller more than the one made by rater with shorter history.

2) **SOM-based** reputation system: In addition to weighting ratings with financial values per-transaction, weights the ratings based on rater's spendings on the market (SOM) as a "proof-of-burn" value. That is, the raters (buyers) are implicitly rated based on how much they spend on this market. So, rating by buyer with a lot of spendings influences reputation more than the one made by rater with smaller spendings.

## 3 Reputation System Evaluation Results

### Case 1. Social Network

Based on our study of social network dynamics [Kolonin, 2019], we were trying to evaluate how the level of reputation computed by WLR algorithm for participants of Steemit social network corresponds to the evaluations of trust and credibility given to these accounts manually. That is, we were computing reputations for entire network for long period of time and comparing the computed reputation ranks with "black lists" maintained by network administrators and volunteers as well as to the lists of "whales", called so for listing well-known publicly available participants. We have studied how the reputation changes over time for accounts of different types, assuming every account starts with default reputation of *0.5* which may get changed to higher or lower in the same very first day and keep changing over time, as it is shown on Figure 2.

The interesting feature of the dynamics is that "expectedly highly reputable" accounts are given longer "tails" spanning over time so reputation either does not decay or decay slower. On the opposite, the "expectedly low reputable accounts" are present with fast reputation value decay. Should be noted, that highly reputable accounts do not necessarily have to get reputation decayed closer to the end of the period – it just have happened that all random accounts selected for the chart were losing reputation to some extent by the end of given time period.

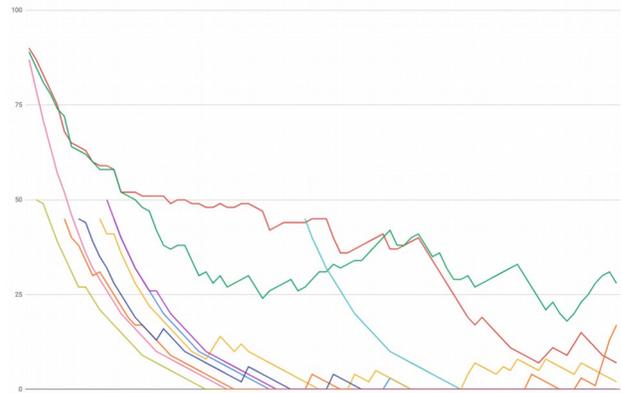

Figure 2. Temporal dynamics of reputation values for randomly selected *5* "expectedly highly reputable" accounts and *5* "expectedly low reputable" ones. Horizontal axis corresponds to time period of *3* months from left to right, vertical axis indicates reputation value in range from *0.0* to *1.0*, labeled on scale from *0* to *100* on the chart.

### Case 2. Marketplace

In our recent work, targeted on study of applicability of the **liquid weighted rank** algorithm for marketplaces, we made a simulations for different populations of 1 hundred, 1 thousand and 10 thousands of suppliers and consumers being active during one quarter, half year and one year, respectively. In the simulations we had 90% of participants as consumers and 10% of them as suppliers. For each of the configurations we have obtained nearly the same results as presented further.

The 5% of all suppliers were dishonest, charging money for fake services and the 5% consumers were dishonest, providing fake positive ratings to the dishonest suppliers at any time. The behavior of the reminding 95% consumers and suppliers was natural so that honest suppliers were providing good real services positively rated by consumers with some fraction of naturally bad quality services negatively rated by consumers. Respectively, the honest consumers encountering fake services served by dishonest suppliers were always rating them negatively and never returning back to such suppliers.

The dishonest suppliers and consumers were acting in campaigns with duration of what we call scam period. When the period ends, all dishonest suppliers and consumers are expectedly changing their previous accounts (identities).

That is, the identities of scammers learned by honest consumers become inactive periodically and the old scammers appear with new identities. That is, multiple dishonest suppliers and consumers appearing in different time on the marketplace can be aliases of real scammers.

We tried four different scam periods: 10, 30, 92 and 182 days. In this way we can compare how the effectiveness of reputation system changes depending on how long scammers stay on the market. It is very hard to say with certainty when they will come on the market and their timing might influence results too much. This is the reason why we decided to make several different simulations in order to compare the performance in different scenarios.

The results can be seen on the following table at Figure 3 for the case of 1 thousand participants active during the half year. We can see that for any scam period, regular reputation system does not provide any improvement, but rather makes things worse, making loss to scam and profit from scam greater. In turn, weighted reputation system provides stable improvement, decreasing loss to scam ad profit from scam. Finally, in cases of TOM-based (assessing raters' reputation with time on the market) and SOM-based (assessing raters' reputation with spendings on market) reputation system with shorter scam periods we have the best improvements.

| Scam Period | Reputation System | Loss to Scam (LTS) | Profit from Scam (PFS) | LTS Relative Decrease | PFS Relative Decrease |
|---|---|---|---|---|---|
| 182 | No | 2.4% | 44% | | |
| 182 | Regular | 2.7% | 49% | -13% | -13% |
| 182 | Weighted | 2.3% | 42% | 2% | 3% |
| 182 | TOM-based | 1.4% | 30% | 41% | 31% |
| 182 | SOM-based | 2.2% | 40% | 8% | 7% |
| 92 | No | 3.0% | 54% | | |
| 92 | Regular | 3.5% | 65% | -19% | -20% |
| 92 | Weighted | 2.8% | 52% | 5% | 4% |
| 92 | TOM-based | 1.7% | 36% | 43% | 33% |
| 92 | SOM-based | 2.6% | 47% | 13% | 12% |
| 30 | No | 3.9% | 73% | | |
| 30 | Regular | 4.7% | 86% | -19% | -18% |
| 30 | Weighted | 3.3% | 59% | 17% | 19% |
| 30 | TOM-based | 1.5% | 31% | 63% | 58% |
| 30 | SOM-based | 1.5% | 27% | 63% | 63% |
| 10 | No | 4.4% | 81% | | |
| 10 | Regular | 4.7% | 88% | -7% | -8% |
| 10 | Weighted | 3.0% | 54% | 33% | 33% |
| 10 | TOM-based | 0.2% | 3% | **96%** | **96%** |
| 10 | SOM-based | 0.3% | 6% | **93%** | **93%** |

**Figure 3.** Comparisons of different configurations of the reputation system algorithm with different periodicity of scam campaigns. The leftmost columns "LTS Relative Decrease" and "PFS Relative Decrease" illustrate the performance of different reputation systems under different scam periods. "LTS Relative Decrease" is the relative increase of loss to scam in relation to having no reputation system. Similarly, "PFS Relative Decrease" is the relative increase of profit from scam in relation to having no reputation system. Regular reputations system corresponds to **page rank**, Weighted reputation system stands for plain **weighted liquid rank (WLR)**, TOM-based and SOM-based correspond to WLR with respective extensions.

In order to evaluate performance of the reputation system, we present the following two metrics of financial kind. **Loss to scam (LTS)**: there we sum up the volume of transactions by honest buyers to dishonest sellers and divide that to the spend of all honest buyer transactions. This metric shows what proportion of money spent by honest buyers is spent on the products offered by dishonest sellers. **Profit from scam (PFS):** we sum up the volume of transactions by honest buyers to dishonest sellers and divide this by spendings of dishonest buyers. This means the return to the money spent by dishonest buyers on their own transactions. It shows how profitable it is to run fraudulent products in our marketplace. This metric should only be understood as relative value, such as what is the improvement of PFS over different systems, since one could argue that dishonest sellers do not need to spend their whole product price to get a fake rating.

If we look at different scam periods, it is clearly seen that if there is no reputation system in use then losses of those agents to scam and profit of scammers increase when the scam period is shortened. Regular reputation system appears rather ineffective for any scam period. Weighted reputation system on the other hand, always shows improvement, increasing with shorter scam periods. Performance of TOM-based and SOM-based systems for longer scam periods of 182 and 92 days is exceeding the one of weighted reputation system, but improves with decrease of scam period significantly, providing the best improvement across all scenarios with the least scam period of 10 days, making losses of honest consumers and profits of scamming suppliers nearly 10 times smaller.

The Figure 3 above shows that using our reputation system based on **weighted liquid rank** algorithm with account to time and spendings on the market, online marketplaces can allocate the products much better and prevent the scam significantly. The biggest winner here are the buyers of the products. They will still lose a bit of money on scamming sellers, however it will be significantly less than they would in case there were no reputation system, as measured with LTS decrease. In a decentralized marketplace with no centralized service for scam prevention, it is still necessary for buyers to spend some money on scam sellers, because at the beginning there is no way of knowing someone is selling good or bad product – only after some feedback from buyers and transactions can we take appropriate action in order to allocate recommendations better.

## Conclusion

Having designed and implemented several generations of the **weighted liquid rank** algorithm extending the **page rank** in application to social systems, we have evaluated performance of the algorithm on real-world and simulation data for social networks and marketplaces. We conclude the solution can applied to these cases improve consensus for democratic governance in these cases and we will be exploring other cases to apply our solution.